\documentclass[aps,amsmath,amssymb,twocolumn,superscriptaddress,showpacs]{revtex4-1}
\usepackage{graphicx}

\begin{document}

\author{King Yan Fong}
\author{Menno Poot}
\author{Hong X. Tang}
\email{hong.tang@yale.edu}
\affiliation{Department of Electrical Engineering, Yale University, New Haven, CT 06511, USA}

\title{Nano-optomechanical resonators in microfluidics}

\begin{abstract}
Operation of nanomechanical devices in water environment has been challenging due to the strong viscous damping that greatly impedes the mechanical motion. Here we demonstrate an optomechanical micro-wheel resonator integrated in microfluidic system that supports low-loss optical resonances at near-visible wavelength with quality factor up to 1.5 million. The device can be operated in self-oscillation mode in air with low threshold power of 45 uW. The very high optical $Q$ allows the observation of the thermal Brownian motion of the mechanical mode in both air and water environment with high signal-to-background ratio. A numerical model is developed to calculate the hydrodynamic effect on the device due to the surrounding water, which agrees well with the experimental results. With its very high resonance frequency (170 MHz) and small loaded mass (75~pg), the present device is estimated to have mass sensitivity of attogram level in liquid environment with bandwidth of 1 Hz.
\end{abstract}

\maketitle

\section{Introduction}
Micro- and nano-scale mechanical resonators have been developed as important tools in both fundamental studies and technological applications. Because of their very small spring constant and mass, even tiny forces acting on the resonator or masses adhering on its surface can greatly alter its dynamics, which can be detected with extremely high precision. Force sensitivity down to sub-attonewton ($<10^{-18}$ N) level has been demonstrated \cite{APL_79_3358_2001}, which allows detection of the force from a single electron spin \cite{Nature_430_329_2004}. As the dimensions of the mechanical devices continue to be scaled down, the mass sensitivity has reached attogram ($10^{-21}~\mathrm{kg}$) \cite{JAP_95_3694_2004, NatNano_2_114_2007}, zeptogram ($10^{-24}$ kg) \cite{NL_6_583_2006}, yoctogram ($10^{-27}$ kg) \cite{NatNano_3_533_2008} level, and down to the mass of a single proton \cite{NatNano_7_301_2012} in the past few years. A recent demonstration showed that this nanomechanical technology holds a promise to perform mass spectrometry on a single molecule with extremely high resolution \cite{NatNano_4_445_2009}.

However, so far most of the high performance nanomechanical sensing systems operate in vacuum or air, only very few operate in liquid due to the rapid deterioration of device performance caused by the enormous fluidic damping. This dramatically lowers the quality factor ($Q$) and the fluid displaced by the movement adds mass. Both effects make operation in liquids extremely challenging. Moreover, the fluidic damping becomes increasingly significant when the resonator dimension is scaled to micro- and nano-scales since the viscous force scales inversely to the square of the characteristic length, i.e., $\mu\nabla^2\vec{v}\sim L^{-2}$. For example, in tapping mode AFM the typical quality factor in liquid are low: $Q < 5$ \cite{APL_64_1738_1994, APL_64_2454_1994}. 

To circumvent the severe damping problem, one approach is to use a hollow resonator structure where fluids and target analytes flow in a fluidic channel embedded inside \cite{Nature_446_1066_2007, NatCommun_4_1994_2013}. The resonator can then be operated in vacuum or ambient air so that the mechanical damping is not compromised. With this approach attogram sensing has been achieved \cite{PNAS_111_1310_2014}. Nevertheless, operation of nanomechanical system in a genuine liquid environment remains important since only then the sensing can take place in environments where many biological and chemical samples naturally reside in \cite{Booksection_Roukes_BioNEMS_2007}. Efforts towards this direction include the development of transduction schemes that can efficiently actuate and readout the motion of the resonator in the highly dissipative liquid environment. Transduction methods such as thermo-optical excitation \cite{NL_6_2109_2006, JAP_99_124904_2006}, magnetomotive drive and detection \cite{APL_95_263103_2009}, and piezoelectric actuation \cite{IEEEIUS_2009_2568}) have been demonstrated. On the other hand, a recent demonstration suggests that the use of higher modes of a cantilever could be a promising candidate for operation in liquid as they generally have higher quality factors and a smaller mass-loading effect from the water \cite{APL_92_043106_2008}.

Besides sensing applications, efficient transduction of nanomechanical resonators in viscous environment also benefits the understanding of fluid dynamics at small length scales. Nanomechanical resonators have been an important tool for exploring new parameter regimes in fluid dynamics, such as the high frequency regime \cite{PRL_98_254505_2007, LabChip_10_3013_2010}, transitions between Newtonian and non-Newtonian regimes \cite{PRL_98_254505_2007, PRL_103_244501_2009} and between hydrodynamic and kinetic regimes \cite{PRL_108_084501_2012}. It also enables the study of stochastic dynamics of fluid-structure interaction due to Brownian noise \cite{PRL_92_235501_2004, Nanotech_17_4502_2006, JAP_84_64_1998}, which can be used for developing highly sensitive AFM in liquid  \cite{JAP_114_144901_2013}.

In this paper, we develop an optomechanical micro-wheel resonator integrated in microfluidic system that operates efficiently in fluidic environment. The resonator supports high $Q$ optical whispering gallery modes, which transduce the mechanical motion with ultra-high sensitivity and allow optomechanical self-oscillation in air with low threshold power. Using the highly sensitive cavity-enhanced optical readout, we are able to resolve the thermomechanical noise of the mechanical radial breathing mode in water with high signal-to-background ratio. To explain the changes in resonance frequency and damping rate caused by the liquid, a boundary integral method is developed to calculate the hydrodynamic function for axial-symmetric systems. Together with its small effective mass (75 pg with the entrained liquid included), high resonance frequency (170 MHz in water), and high displacement sensitivity (15 $\mathrm{am/\sqrt{Hz}}$), mass sensitivity of 2.5 attogram with measurement bandwidth of 1 Hz is predicted. This device shows promise in sensing applications in fluidic environment. Our design of the integrated system and the robust fabrication process also provide a viable approach for large scale integration of nanophotonic devices with microfluidics.

\section{Device design and fabrication}

\begin{figure*}[!t]
\centering
\includegraphics[width=0.75\textwidth]{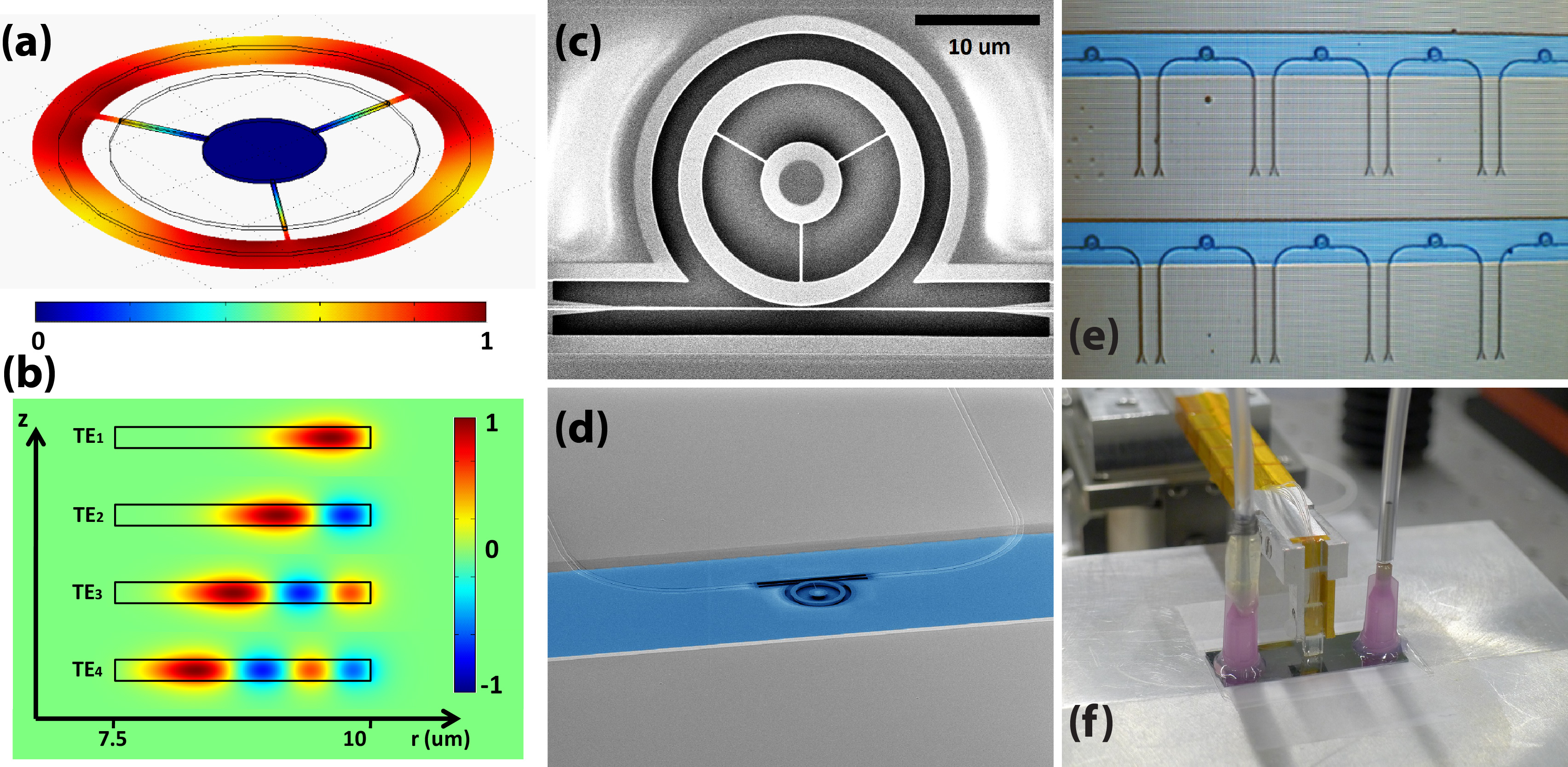}
\caption{ (a) Normalized displacement profile of the fundamental radial breathing mode. (b) Electric field (radial component) distribution of the first four TE modes ($\mathrm{TE}_{i}$) in air, where the index $i$ represents the mode orders in the radial directions. (c) Scanning-electron microscope image showing a top view of the device. (d) Angled view SEM image showing the device positioned in the middle of the microfluidic channel (highlighted in blue) (e) Optical top view of the chip showing arrays of devices and two microfluidic channels (highlighted in blue). (f) Photo of a fully packaged device aligned to a fiber probe. } \label{fig1}
\end{figure*}

One of the technical obstacles of realizing photonic resonators in water environment is the optical absorption due to the water, which can severely degrade the optical $Q$ and cause undesirable thermo-optical heating of the surrounding water. It is well known that water is strongly absorptive in the near-infrared range where most integrated optomechanics experiments are done. The optical extinction coefficient at wavelength of $\lambda=1.55~\mathrm{\mu m}$ is $\kappa=9.86\times 10^{-5}$, \cite{*[{Interpolation from the data from }] [{}] AO_12_555_1973} which corresponds to a loss of -34 dB/cm for an electromagnetic plane wave propagating in the medium. Therefore, telecommunication wavelength commonly used in silicon photonics is not suitable for this purpose. Although 1D photonic crystal resonators at this wavelength range has been demonstrated in water, a low quality factor $Q\approx 3000$ was obtained \cite{LabChip_9_2924_2009}. To achieve high $Q$ optical resonances in water, here we developed a photonic resonator that operates in near-visible wavelength at around 780 nm. The extinction coefficient of water at this wavelength is about 700 times smaller: $\kappa=1.43\times 10^{-7}$. \cite{*[{By Interpolation of the data from }] [{}] AO_12_555_1973} This is equivalent to an attenuation of only -0.1 dB/cm, which would result in an absorption-limited $Q$ of 10 million. This wavelength also falls within the biological transparency window
\cite{NatBiotech_19_316_2001} and is therefore useful in biology and for medical applications. For the waveguiding material, stoichiometric silicon nitride is used because of its low optical absorption in the visible range, as shown by a recent demonstration of optical resonances with $Q$s up to 3 million in the 652 -- 660 nm wavelength range. \cite{OE_17_14543_2009}

The design of the optomechanical resonator in this study is a suspended micro-wheel structure \cite{IEEEMEMS_2004_821, OE_19_22316_2011, OE_19_24522_2011, APL_100_171111_2012}. Mechanically it supports a radial breathing mode, which has been demonstrated to have high mechanical quality factor up to $Q_M\approx4000$ at 1.3 GHz operated in air \cite{OE_19_22316_2011}. The motion of the radial breathing mode does not involve change of center of mass. Therefore, it is expected to cause minimum energy loss due to radiation (i.e., mass transport) through the fluidic environment. Fig. \ref{fig1} (a) shows the displacement profile of the fundamental radial breathing mode of a micro-wheel of 10 um radius simulated using finite element analysis. This mode has an effective mass of 65 pg and a resonance frequency of 175 MHz in vacuum.

The mechanical mode is optomechanically coupled to the optical whispering gallery modes \cite{OE_15_17172_2007}: as the displacement of the mechanical mode vibrates in the radial direction, it changes the resonance wavelength of the optical mode, which can be used as the readout of the mechanical motion. Fig.~\ref{fig1}~(b) shows a cross-sectional plot of the simulated radial electric field distribution of the first four transverse-electric (TE) modes in air. The optical mode is mainly confined at the outer rim of the micro-wheel. Therefore scattering loss due to the presence of the anchoring spokes at the inner rim has a negligible effect. For the case when the device is immersed in water, simulation shows that the fourth TE mode becomes not well confined due to the smaller index contrast.

The device is fabricated in a 200 nm thick stoichiometric silicon nitride film deposited by low-pressure chemical vapor deposition (LPCVD). Details of the fabrication procedures can be found in the supplementary material. The device layer is cladded with a 2 um thick thermal oxide from the bottom and a 3 um thick plasma-enhanced chemical-vapor deposited (PECVD) oxide from the top. Microfluidic channels were crafted into the top cladding layer directly above the device exposing the resonator structure to the external environment. The chip was bonded to a thin cover glass slide using PDMS as a bonding layer. For each device, an evanescently coupled waveguide is fabricated next to the resonator and a pair of gratings is used for coupling light into and out of the waveguide. Fig. \ref{fig1} (c) and (d) show SEM images of a device with outer radius of 10 um and ring width of 2.5 um before the cover glass bonding. As can be seen in the figure, a tapered structure is designed to provide a robust support to the free-standing waveguide. In fact, the whole structure is sturdy enough that it can be released directly in a wet undercut process without the use of a critical point dryer. This allows repeatedly switching between operation in liquid and gaseous environments. Fig. \ref{fig1} (d) shows a device positioned in the middle of a microfluidic channel, which is highlighted in blue color. Fig. \ref{fig1} (e) is an optical image showing the top-view of the chip after the glass bonding. A fully packaged device is shown in Fig. \ref{fig1} (f), where a fiber array probe is approached from the top and aligned to the grating couplers on the chip. This design seamlessly integrates the microfluidic and photonic systems and allows efficient fiber access to the on-chip waveguides. It provides a viable approach for large scale integration of nanophotonic devices with microfluidic system.

\begin{figure}[t]
\centering
\includegraphics[width=7cm]{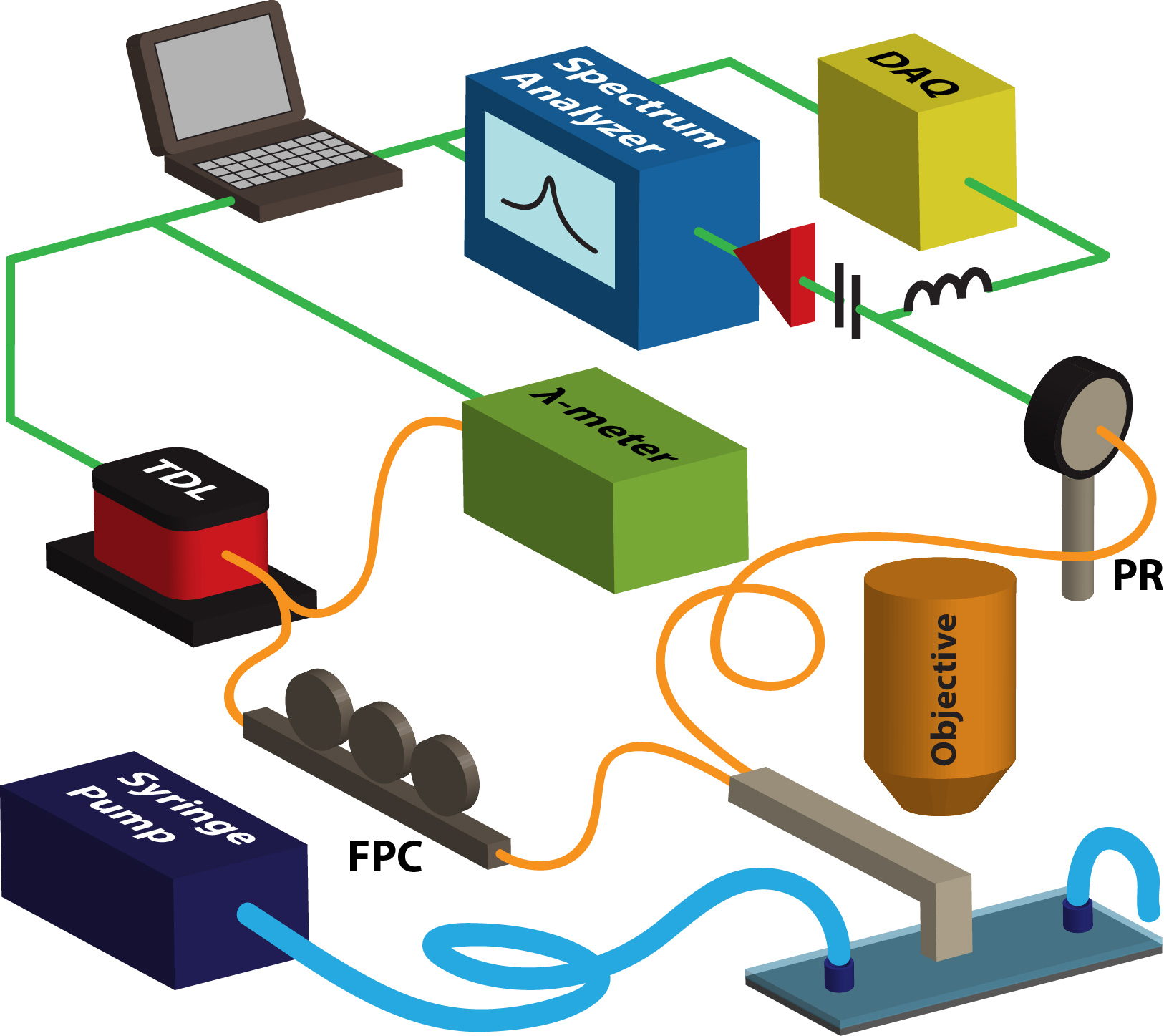}
\caption{ (a) Schematic of the measurement setup. TDL: Tunable diode laser. FPC: Fiber polarization controller. PR: Photoreceiver. DAQ: Data acquisition system. }
\label{fig2}
\end{figure}

The device was characterized using the setup shown in Fig. \ref{fig2}. A tunable diode laser (New Focus TLB-6712) with a tuning range of 765 -- 781 nm was used to probe the optical spectral response of the device. A small portion of the laser output was tapped out and sent to a wavelength meter for wavelength and intensity calibration. Using a fiber polarization controller, the laser light was adjusted to be TE-polarized to optimize the coupling efficiency of the grating couplers. Next, a fiber probe was aligned from the top to the grating couplers on the device and the transmitted light was collected on a high speed photoreceiver (New Focus 1601, 1 GHz bandwidth). Part of the signal was sent to a data acquisition system to measure the dc transmission while the other part was sent to an electrical spectrum analyzer for spectral characterization. In this setup, all the optical fibers used, including the ones in the fiber array, are single mode fibers (SM800, core size of 5.6 um) designed for the wavelength range at around 780 nm.

\begin{figure*}[!t]
\centering
\includegraphics[width=0.7\textwidth]{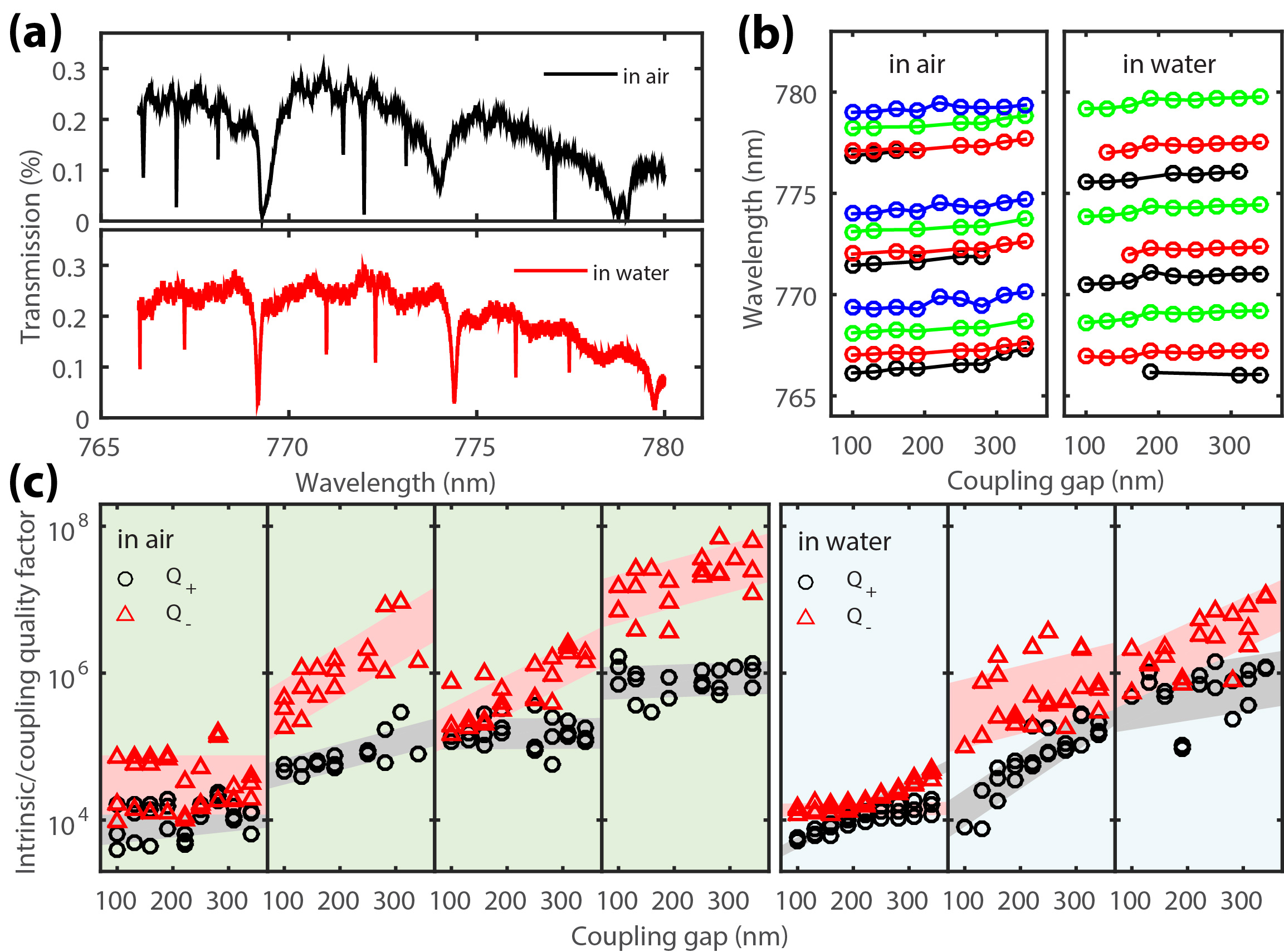}
\caption{ (a) Optical transmission spectra taken in air and water, with coupling gaps of 100 nm and 310 nm respectively. (b) Resonance wavelengths of the devices plotted against coupling gap. Resonances belong to the same radial mode are plotted in same color. (c) Quality factors $Q_\pm$ of different radial modes plotted against coupling gap. } \label{fig3}
\end{figure*}

\section{Experimental results}

The measured optical transmission spectra of the device (outer radius of 10 um and ring width of 2.5 um) in air and water, which includes the insertion loss of the two input-output grating couplers, are shown in Fig. \ref{fig3} (a). The peak transmission reaches $\sim 0.3\%$, corresponding to an insertion loss of $-13$ dB per grating coupler. The coupling efficiency is limited by the separation between the fiber probe and the grating couplers set by the thickness of the cover glass, which is around 100 um. In the measured spectra, groups of resonances having similar quality factors and extinction ratios can be identified. They correspond to the TE whispering gallery modes of different radial orders. The resonance wavelengths are plotted against the coupling gap $g$ in Fig. \ref{fig3} (b). Resonances of the same radial mode are plotted in the same color. Four modes can be identified in the spectra taken in air while three modes show up in case of water, which agrees with the simulation that the fourth TE mode in water is not well confined.

Different radial modes show distinct behavior in quality factor and extinction ratio. The loaded quality factor $Q$ and the normalized on-resonance transmission $T_0=P_{out}(\lambda=\lambda_0)/P_{in}$ are related to the intrinsic quality factor $Q_i$ and coupling quality factor $Q_c$ by
\begin{eqnarray}\label{eq:extc_QL}
Q^{-1} &=& Q_i^{-1} + Q_c^{-1} \label{eq:QL}\\
T_0 &=& \left(\frac{Q_c-Qi}{Qc+Qi}\right)^2 ~.\label{eq:T0}
\end{eqnarray}
From the measured $Q$ and $T_0$, $Q_i$ and $Q_c$ can be calculated, i.e. $Q_c,Q_i=Q_\pm=2Q/(1\pm\sqrt{T_0})$. From this it follows that $Q_- > Q_+$ but it cannot be determined from Eqs. (\ref{eq:QL}) and (\ref{eq:T0}) which one corresponds to $Q_c$ and which to $Q_i$. This depends on whether the system is over-coupled ($Q_c<Q_i$) or undercoupled ($Q_c>Q_i$). One way to resolve this ambiguity is to observe how $Q_+$ and $Q_-$ vary with the coupling gap $g$, since $Q_c$ is expected to be an increasing function of $g$ while $Q_i$ is expected to be independent of $g$. Fig. \ref{fig3} (c) plots the measured $Q_\pm$ against the coupling gap for different radial modes. For each radial mode, we attribute the $Q_\pm$ that has a weaker (stronger) dependence on $g$ as the intrinsic (coupling) quality factor $Q_i$ ($Q_c$). Note that while $Q_i$ is expected to show no dependence on $g$, some of them do decay slowly as $g$ decreases, which could be due to the increasing scattering loss in the proximity of the coupling waveguide. Also note that for the first mode in water, $Q_+$ and $Q_-$ display a crossing feature: the system changes from the undercoupled regime at large $g$ to being overcoupled at small $g$. Among all the measured results, the highest loaded quality factor obtained is $(1.53\pm 0.04)\times10^6$ in air and $(1.50\pm 0.03)\times10^6$ in water. It is thus indeed possible to maintain the high $Q$ that was obtained in air when operating the device in water.
\begin{figure}[t]
\centering
\includegraphics[width=8cm]{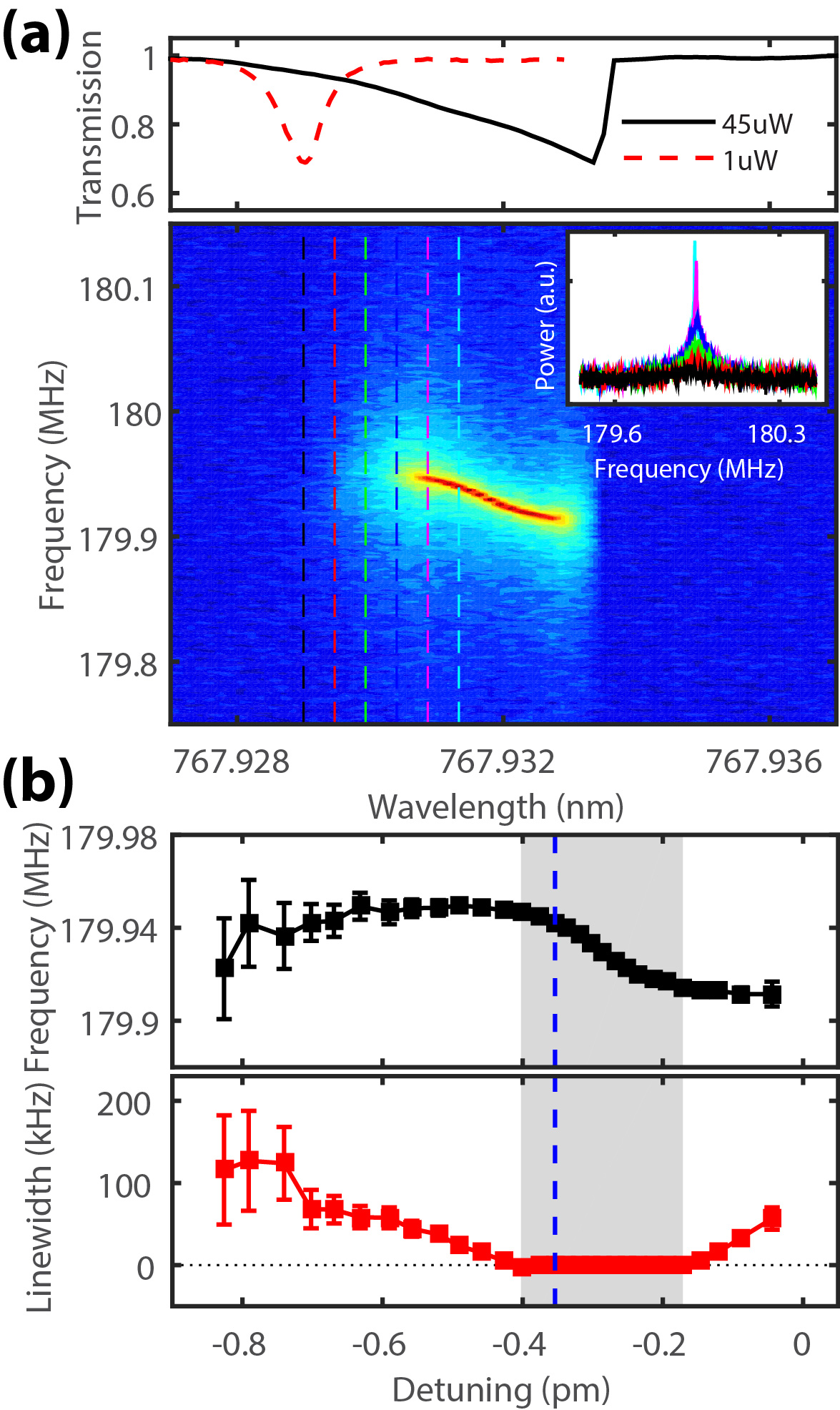}
\caption{ Optomechanical oscillations in air. (a) Upper panel shows the normalized transmission across a resonance at different laser power. Lower panel shows the contour plot of the measured RF spectrum versus wavelength when the laser power is 45 uW (corresponds to the black line in the upper panel). Cross-sectional plots of the spectra at different wavelengths indicated by the dashed-lines are shown in the inset. (b) Mechanical frequency and linewidth plotted against detuning from the optical resonance. The region where the device undergoes self-sustained oscillations is shaded in grey color. Blue dashed-line indicates the detuning that corresponds to the half-linewidth of the optical resonance. Black dotted-line shows the level of zero linewidth. } \label{fig4}
\end{figure}

For an optomechanical device with such high optical $Q$, the optomechanical dynamical backaction can have a significant effect on the mechanical motion; it can either amplify or dampen the mechanical motion depending on the laser detuning \cite{OE_15_17172_2007}. When the input laser power is high enough, the amplification effect can completely compensate the intrinsic mechanical damping and cause the resonator to self-oscillate, which has been shown to be useful in sensing application \cite{OE_21_19555_2013}. Because of the very high optical $Q$, the present device can be operated in self-oscillation mode in air with low threshold power. Fig.~\ref{fig4}~(a) shows the normalized transmission and the RF spectrum when the wavelength is scanned across the resonance with an input power of 45 uW. The transmission curve shows an asymmetric shape caused by the thermo-optical bistability \cite{OL_29_2387_2004} where the heating due to the optical energy in the cavity moves the resonance to longer wavelengths. The transmission at even lower power (1 uW) is shown in the same figure for comparison between the linear and nonlinear regimes. As the laser wavelength approaches from the shorter wavelength side of the resonance, the mechanical resonance peak becomes sharper and higher, indicating the effect of optomechanical amplification. Assuming the optical linewidth remains the same throughout the wavelength scan, the actual detuning can be inferred from the transmission value and the original linewidth. The mechanical frequency and the linewidth at different detuning is plotted in Fig.~\ref{fig4}~(b). As expected, when the detuning approaches half the linewidth of the cavity (indicated by the dashed blue line), the mechanical linewidth decreases until it reaches zero, where the device undergoes large amplitude self-sustained oscillation (region shaded in gray color in the figure) \cite{PRA_86_053826_2012}. As the detuning is further reduced, the optomechanical effect diminishes and the mechanical linewidth gradually return to its original values until the thermo-optical bistable point is reached.

\begin{figure}[t]
\centering
\includegraphics[width=8cm]{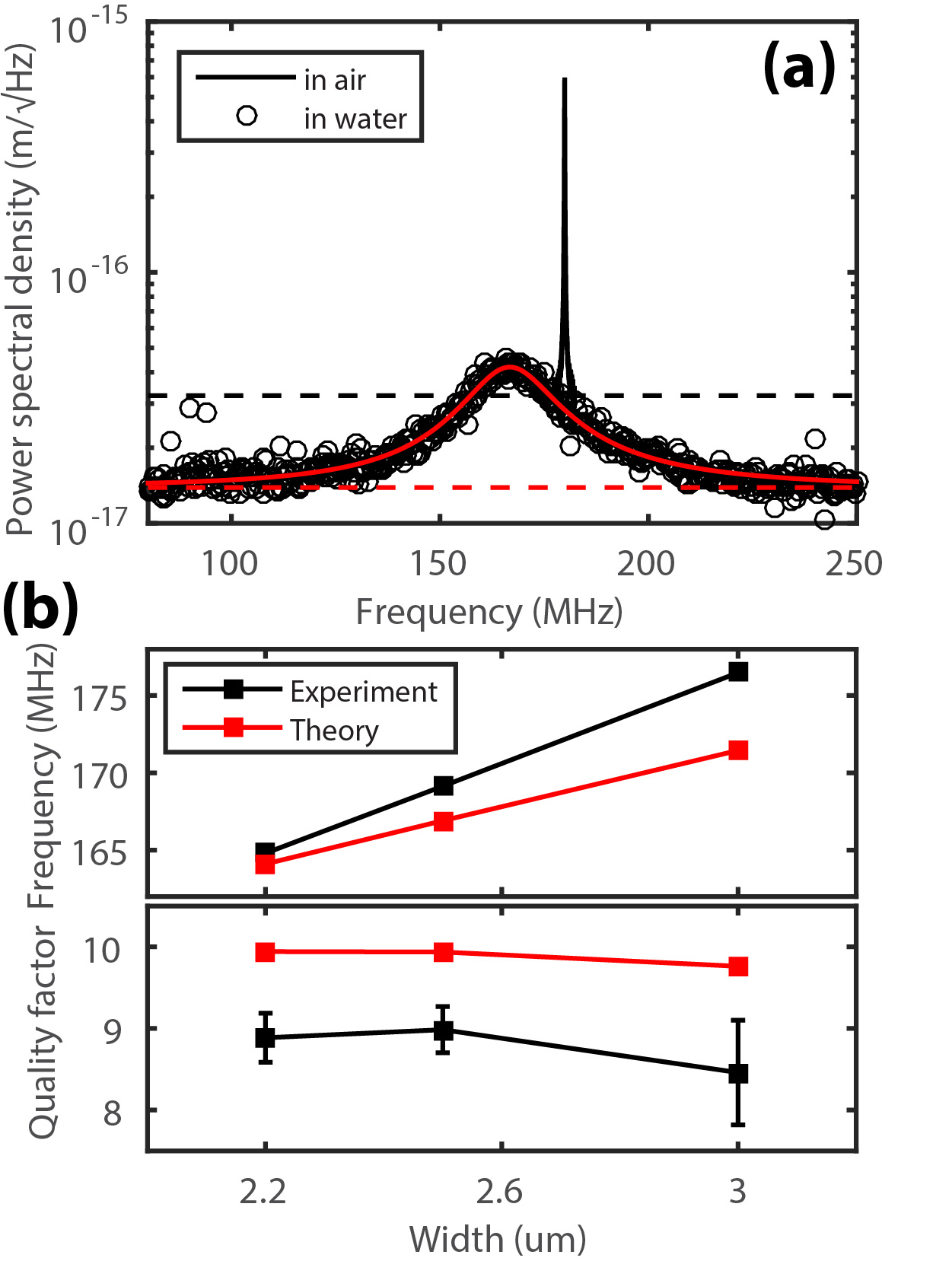}
\caption{ (a) Thermomechanical noise spectrum of the radial breathing mode in air and water environment. Red solid line represents the fitting to the water data. Black and red dashed lines indicate the noise floors of the measurement in air and water respectively. (b) Resonance frequency (upper panel) and quality factor (lower panel) plotted against the width of the micro-wheel. $r_{out}$ is fixed as 10 um. The experimental (numerical) results are shown in black (red). }
\label{fig5}
\end{figure}

When the device is immersed in water, the viscous damping is so great that we were not able to set the device into self-sustained optomechanical oscillation. Nevertheless, with the very high optical $Q$, the thermomechanical noise of the mechanical mode can be resolved even when the mechanical motion is highly dampened in water. Fig. \ref{fig5} (a) shows the thermomechanical noise spectra of the mechanical resonance measured in both air and water environment. In both cases the thermal motion is visible as a peak on top of the detector imprecision noise. In air, the peak is narrow and high; there the radial breathing mode has a resonance frequency of 179.9 MHz and mechanical quality factor of $Q_M=2160$. In water, the frequency is slightly shifted down to 169.4 MHz and $Q_M$ is reduced to 9. Note that the noise floors of the spectra taken in air and water are different, as indicated by the dashed lines in the figure, because a lower optical power was used in air to minimize the effect of the optomechanical backaction. For the noise spectrum in water, a noise floor of $S_{nn}^{1/2} = 15~\mathrm{am/\sqrt{Hz}}$ is achieved.

It is noteworthy that the resonance frequency drops by only 5.8\% after immersing the device in water. This is very small compared to other demonstrated mechanical systems. For example, AFM cantilevers working at several hundreds of kHz 
can have resonance frequency that drop to 1/2 -- 1/5 of the original (air) value when immersing in water
\cite{RSI_67_3583_1996}. 
Also silicon nitride nanostring resonators working at 100 MHz range show a more than 25\% drop in resonance frequency \cite{NL_6_2109_2006}. Another observation is that, the quality factor of the present device is also higher than other demonstrated systems, which typically have $Q_M$ of 5 or lower \cite{APL_64_1738_1994, NL_6_2109_2006, JAP_99_124904_2006, APL_95_263103_2009, IEEEIUS_2009_2568, RSI_67_3583_1996}. The smaller frequency shift and higher $Q_M$ hints that the hydrodynamic mass loading and damping due to the surrounding water is relatively small for the micro-wheel resonator structure.

The dynamics of the micro-wheel resonator inside a liquid environment can be modeled with the equation of motion for a single harmonic oscillator after integrating over the mode shape:
\begin{equation}\label{eq:EOM}
\ddot{x}(t) +2\gamma\dot{x}(t) +\Omega^2x(t) = \frac{F_{f}(t)}{m}+\frac{F_{th}(t)}{m} ~,
\end{equation}
where $\Omega$, $\gamma$ and $m$ are the resonant frequency, damping rate and effective mass in vacuum, and $F_{f}(t)$ and $F_{th}(t)$ are the fluidic force and thermal fluctuation force acting on the resonator. In the limit of small oscillation amplitude, the fluidic force can be assumed as a linear response function of the resonator motion, i.e., $F_{f}(t)=\int dt^\prime\chi(t-t^\prime)\dot{x}(t^\prime)$. In the frequency domain, it can be written as $F_{f}[\omega]=m\omega^2 \Gamma[\omega]x[\omega]$, where $\Gamma[\omega]= -i\chi[\omega]/m\omega$ is a dimensionless function called the ``hydrodynamic function''. The definition adopted here differs from those in Refs. \cite{JAP_84_64_1998, PRL_92_235501_2004} by a geometrical factor. In general $\Gamma[\omega]$ is a complex function. Its real part $\Gamma_R[\omega]$ and imaginary part $\Gamma_I[\omega]$ are related to the mass loading and damping respectively due to the presence of the fluid. Using the fluctuation-dissipation theorem, it can be shown that the single-sided noise power spectral density of the displacement due to thermomechanical fluctuation is given by \cite{PRL_92_235501_2004, Nanotech_17_4502_2006}
\begin{equation}\label{eq:thermo_spec}
\mathcal{S}_{xx}[\omega]= \frac{(4k_BT/m\Omega^3) \tilde{\omega}\Gamma_I[\omega]} {(1-\tilde{\omega}^2(1+\Gamma_R[\omega]))^2 +(\tilde{\omega}^2\Gamma_I[\omega])^2} ~,
\end{equation}
where $\tilde{\omega}=\omega/\Omega$ is the normalized frequency. Here the intrinsic damping rate $\gamma$ is assumed to be negligible compared with the hydrodynamic damping, i.e. $\gamma\ll\omega\Gamma_I[\omega]$.

For a cantilever with rectangular geometry, the hydrodynamic function can be solved numerically using the boundary integral method described in Refs. \cite{JEngMath_3_29_1969, JAP_84_64_1998, PhysFluids_22_052001_2010}. A formula introduced in Ref. \cite{JAP_84_64_1998} provides a good approximation for the hydrodynamic function of a cantilever with large width-to-thickness ratio. These studies are relevant to mechanical resonators with rectangular geometry such as AFM cantilevers. To study the hydrodynamic effect on a micro-wheel resonator, here we use a boundary integral method for systems with rotational symmetry and solve the corresponding Green's equation. The details of this model can be found in the supplementary material.

From the hydrodynamic function, the loaded resonance frequency $\Omega_w=\Omega(1+\Gamma_R[\Omega_w])^{-1/2}$ and quality factor $Q_w=(1+\Gamma_R[\Omega_w])/\Gamma_I[\Omega_w]$ in water can be calculated \cite{JAP_84_64_1998}. Fig. \ref{fig5} (b) plots the resonance frequency and quality factor as function of ring width for devices with fixed outer radius $r_{out}=10$ um. The symbols are mean values and the error bars are standard deviations for four devices per data point. As shown in the figure, the numerically calculated $Q_M$ is around 10 while the measured $Q_M$ is around 8.5 to 9, which agree fairly well with each other. The numerical mode also predicts the observed width dependence.

The total loaded mass of the device including the water entrained to the resonator motion is given by $m_w=m(1+\Gamma_R[\Omega_w])=75$ pg, which is roughly equal to the sum of the original (effective) mass of 65 pg and the mass of water enclosed in the Stokes boundary layer around the device (10 pg). The Stokes boundary layer is a measure of how far the fluid oscillations extend from the device and has a thickness $\delta=31$ nm (See supplementary materials). We would like to emphasize that the entrained mass of water is only of a small fraction (15\%) of the original resonator mass. This is orders of magnitude smaller than that of a typical cantilever structure, where the added water mass can be 10s of times higher than the resonator mass \cite{RSI_67_3583_1996, APL_92_043106_2008}. Even for the very lightweight silicon nitride nanostring resonator, the added water mass is more than 1.3 times the original mass \cite{NL_6_2109_2006}. The small water entrainment of our device is the result of the small resonator dimensions, high resonance frequency and thus small Stokes boundary layer thickness, and in-plane motion of the radial breathing mode, which is expected to have smaller mass loading effect in liquid compared to out-of-plane motion \cite{PhysFluids_22_052001_2010}, 

The values of both the resonator mass and entrained water mass are low, indicating that our device will have a large frequency shift $\Delta f = (f_w/2m_w)\Delta m$ for a small added mass $\Delta m$. This frequency shift is measured by driving the resonator to a coherent oscillation with amplitude $x_0$ and detecting the phase between the drive and the motion. The phase noise in such a measurement determines the mass sensitivity of the device $\delta m$. The phase noise contains both the imprecision noise ($S_{nn}$) and the resonator's thermal motion ($S_xx$). Both contributions can be estimated from the measured noise spectrum (cf. \ref{fig5}~(a)). The mass sensitivity of the resonator is given by
\begin{equation}\label{eq:mass_sensitivity}
\delta m=\frac{m_{w}}{Q_w x_0} \left[ \frac{1}{2} \left(\mathcal{S}_{xx}[\Omega_w] +S_{nn}\right) \frac{\Delta\Omega}{2\pi} \right]^{\frac{1}{2}}~,
\end{equation}
where the measurement bandwidth has been assumed to be much less than the mechanical linewidth, i.e., $\Delta \Omega \ll \Omega_w/Q_w$. The factor of $1/2$ inside the square root accounts for the fact that only half of the noise power is distributed in the phase quadrature. From Fig.~\ref{fig5}~(a), the sum of the thermomechanical noise and the measurement noise background gives a noise peak of $42~\mathrm{am/\sqrt{Hz}}$ on resonance. Assuming the device can be driven to an amplitude of $x_0=100$ pm (mechanical strain of $10^{-5}$) before reaching the mechanical nonlinearity limit, the mass sensitivity is estimated to be $\delta m / (\Delta \Omega / 2\pi) = 2.5 ~\mathrm{ag/\sqrt{Hz}}$. This high mass sensitivity in water environment comes as a result of small mass loading and high quality factor of the device. So far, attogram sensing in fluid has only been achieved using cantilevers with embedded channels \cite{PNAS_111_1310_2014}. Our results suggest that optomechanical micro-wheel resonator can be a promising candidate for in-situ attogram sensing in water environment.

\section*{Acknowledgements}
M.P. thanks the Netherlands Organization for Scientific Research (NWO)/Marie Curie Cofund Action for support via a Rubicon fellowship. H.X.T. acknowledges support from a Packard Fellowship in Science and Engineering and a career award from National Science Foundation. This work was funded by the DARPA/MTO ORCHID program through a grant from the Air Force Office of Scientific Research (AFOSR).

\bibliography{../../../../References/library}

\clearpage

\section{Supplementary materials}

\subsection{Device fabrication}

\begin{figure}[!b]
\centering
\includegraphics[width=8.5cm]{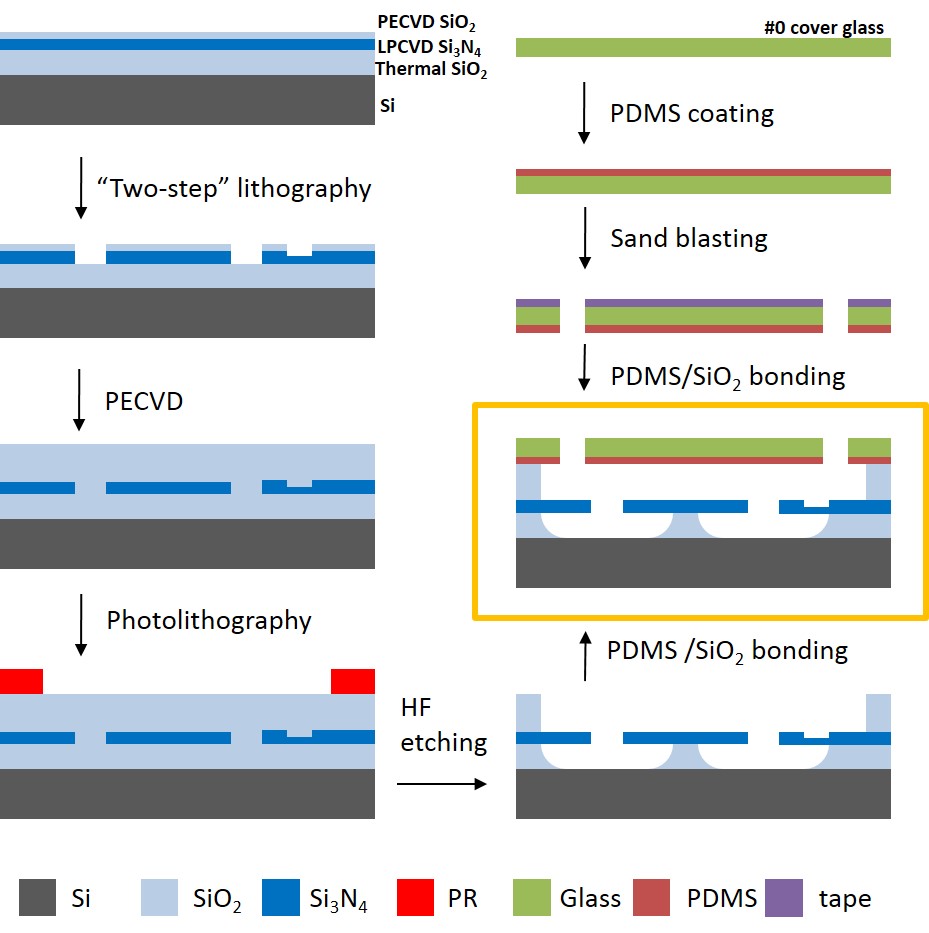}
\caption{ Fabrication process flow of the optomechanical micro-wheel resonator. The final result is highlighted by the orange box. }
\label{fig:Supp_Fab}
\end{figure}

This section presents the details of the fabrication process for the optomechanical micro-wheel resonator. A schematic illustration of the process flow is shown in Fig.~\ref{fig:Supp_Fab}. The device is patterned using the two-steps e-beam lithography and dry etching technique employed previously by our group \cite{APL_100_171111_2012}. The process starts with a silicon wafer with 2 um thermal oxide, 200 nm LPCVD stoichiometric silicon nitride, and a 100 nm thick PECVD silicon dioxide. First, e-beam lithography is performed to define the device pattern in an e-beam resist (ZEP520A). After developing, the e-beam resist is reflowed at 150 $\mathrm{^oC}$ for 1.5 min to smoothen the sidewall. Reactive-ion etching (RIE) with fluorine chemistry is then used to dry etch the top PECVD oxide and half (100 nm) of the silicon nitride layer. The e-beam resist is removed by $\mathrm{O_2}$ plasma afterward and another ZEP520A resist is spin-coated on the chip for the second e-beam lithography. The second e-beam lithography defines the region where the bottom silicon oxide is to be exposed to HF solution for releasing. After the second lithography, another RIE is carried out to etch away the remaining half of the silicon nitride layer. During the second RIE, the top PECVD oxide functions as masking layer to protect the part of photonic structure that is exposed to the plasma making the two patterns self aligned. Again, the e-beam resist is removed by $\mathrm{O_2}$ plasma and the chip is further cleaned in piranha solution.

Next, 3 um thick PECVD silicon dioxide is deposited on top of the chip. The chip is again cleaned in piranha solution and thoroughly dehydrated in an oven. It is then primed with HDMS and spin-coated with photoresist (Shipley S1813). Photolithography is performed to pattern the microfluidic channels. The chip is post-baked at 120 $\mathrm{^\circ C}$ for 10 min to further harden the resist. The chip is then immersed in a buffered HF solution. In this way, the etching of the microfluidic channels and the releasing of the micro-wheel is done in one step. Finally, the chip is transferred to isopropanol and flash dried on hotplate. At this stage, we have a chip of optomechanical devices located inside microfluidic channels (See Fig. \ref{fig:Supp_Fab} bottom right, and Fig. \ref{fig1} (d) in the main text).

The next step is to seal the microfluidic channels by bonding the chip to a cover glass slide. In order to minimize the distance between the fiber array and the on-chip grating couplers, a size \#0 cover glass (80 - 100 um thick) is used, which is the thinnest among the standard glass slides. To bond the cover glass to the PECVD dioxide layer of the chip, a thin PDMS coating is used for adhesion. PDMS is prepared by mixing the PDMS elastomer and the curing agent in 10:1 ratio. The mixture is then diluted in hexane with 1:1 vol. ratio and is thoroughly mixed using a vortex mixer. The diluted PDMS solution is spin-coated on a pre-cleaned cover glass at 3000 rpm for 30 s and is cured overnight in an oven at 100 $\mathrm{^\circ C}$. This results in a 300 nm thick film. After the PDMS is completely cured, an adhesive tape punched with holes (diameter of 2 mm) was taped on the glass side to act as the mask for the subsequent sand blasting process. Sand blasting with $\mathrm{Al_2O_3}$ powder is used to drill holes in the cover glass. The dust generated during the process can be removed using adhesive tape. To proceed to the bonding, the glass piece and the chip are put in $\mathrm{O_2}$ plasma with an RF power of 50 W for 30 s. The two pieces are then aligned and pressed against each other. The bonded chip is put in the oven for overnight baking. For the final packaging, tubing adapters are glued to the cover glass at the hole area using epoxy. Tubings are attached to the adapter for liquid transfer.

\subsection{Wavelength calibration}

\begin{figure}[!t]
\centering
\includegraphics[width=8cm]{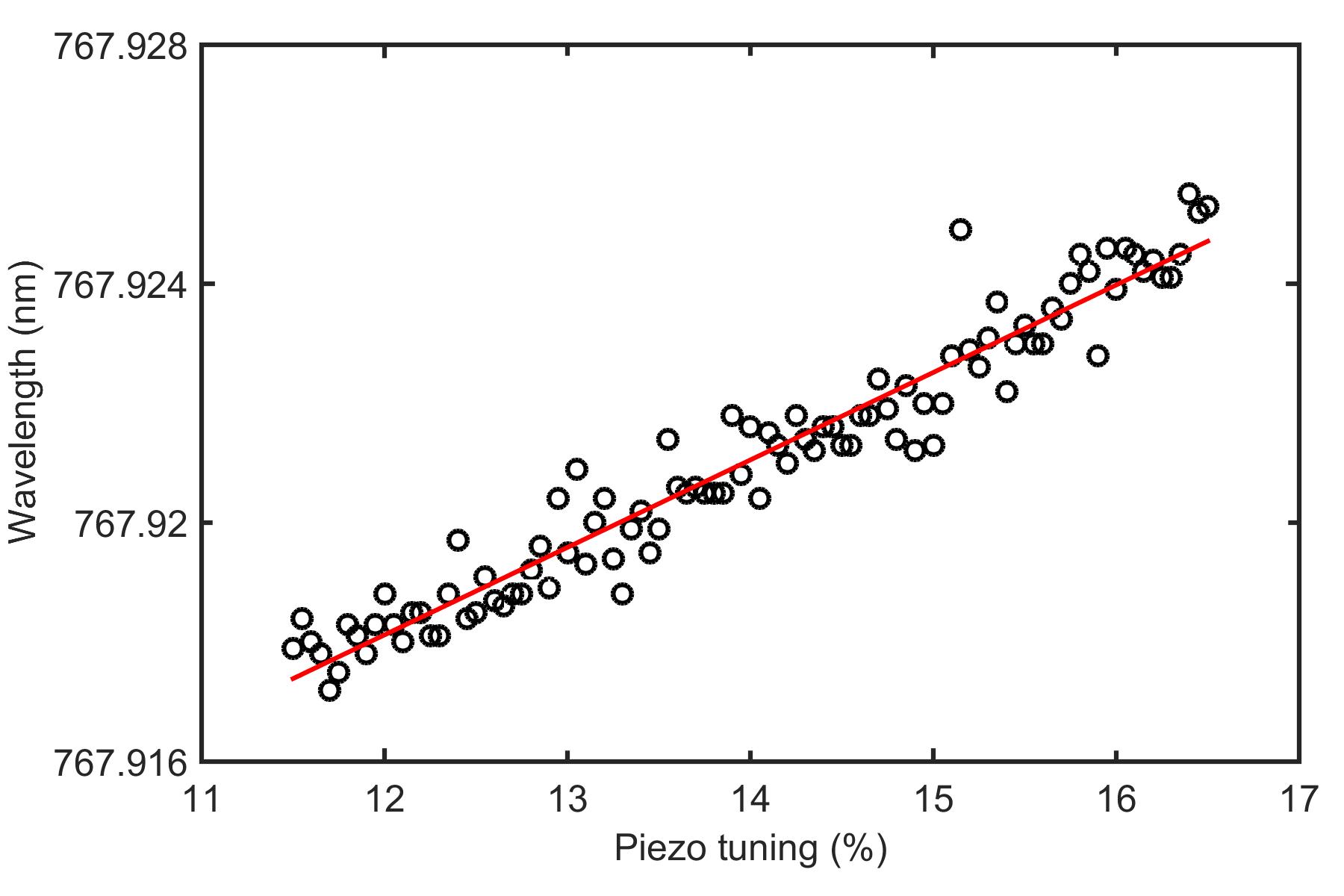}
\caption{ Laser wavelength plotted as a function of the piezo voltage (expressed as a percentage of the full range). }
\label{fig:Supp_wave_cal}
\end{figure}

To quantify ultra-high $Q$ optical resonances, precise calibration of the laser wavelength is very important. The tunable diode laser used in the measurement (New Focus TLB-6712) has a piezo tuning mechanism to fine tune the cavity length and thus the laser wavelength. To calibrate the exact wavelength shift induced by the voltage applied to the piezo, we use a wavelength meter (HP86120) to measure the laser wavelength. Fig. \ref{fig:Supp_wave_cal} shows a calibration curve of the laser wavelength as the piezo voltage is varied. Fluctuation in the wavelength readings is due to instrument noise, which depends on the measurement bandwidth. The result shows that the laser wavelength can be tuned with a high degree of linearity and precision.

\subsection{Numerical model of hydrodynamical loading}
This section describes the theoretical model for calculating the hydrodynamic function of systems with rotational symmetry. The hydrodynamic effects on an oscillating solid structure can be understood using the following physical picture. As the boundary of the structure oscillates with a certain amplitude $A$, the surrounding fluid is displaced and the motion propagates through the medium with dynamics governed by the Navier-Stokes equations. Eventually the whole system reaches a steady state. The fluid force acting on the solid structure can be calculated from the pressure field and the viscous dragging of the fluid. When the oscillation amplitude is small, the fluidic force is expected to be linear in $x(t)$ and, in general, to have a relative phase shift compared to the mechanical motion. The phase shift means that part of the force is in-phase with the velocity which causes damping, and part of the force is in-phase with the displacement which causes mass loading. The above scenario can be formulated as a partial differential equation problem of solving Navier-Stokes equation. For the boundary condition at the fluid-structure interface, the no-slipping condition is assumed, which means that the velocity of the fluid at the surface follows the velocity of the mechanical motion of the solid structure exactly. We also assume the fluid to be incompressibile, i.e., $\nabla\cdot\vec{v}=0$, where $\vec{v}$ is the velocity field of the fluid.

\begin{figure}[!t]
\centering
\includegraphics[width=7cm]{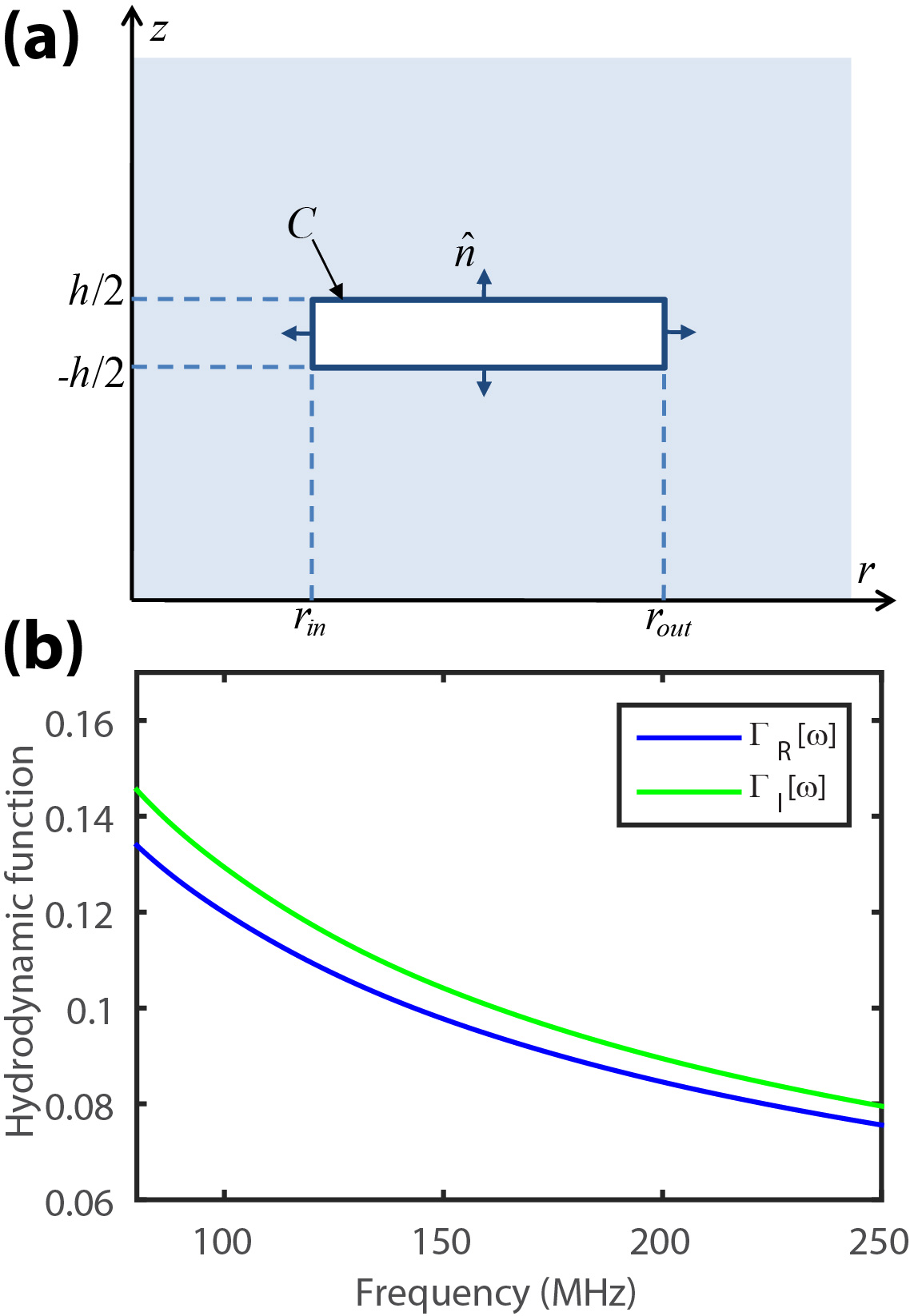}
\caption{ (a) Cross-sectional view of the micro-wheel structure in cylindrical coordinates. (b) Real and imaginary parts of the calculated hydrodynamic function plotted against frequency. }
\label{fig:Supp_model}
\end{figure}

Reynold's number is an important scaling parameter in fluid dynamics. In general, one can define two Reynold's numbers based on the ratios between the inertial force $\rho\frac{D}{Dt}\vec{v} =\rho\frac{\partial}{\partial t}\vec{v} +\rho(\vec{v}\cdot\nabla)\vec{v}$ and the viscous force $\mu\nabla^2\vec{v}$. For oscillatory motion, the time-dependent term of the inertial force gives rise to the frequency-based Reynold's number $\mathrm{Re}_f=\rho\omega L^2/\mu$ and the nonlinear term of the inertial force gives rise to the velocity-based Reynold's number $\mathrm{Re}_u=\rho\omega AL/\mu$ \cite{Booksection_Roukes_BioNEMS_2007}, where $\rho$ and $\mu$ are the density and dynamic viscosity of the fluid, $L$ is the characteristic length scale of the system, and $A$ is the oscillation amplitude.  One can immediately see that for small-amplitude oscillations, i.e., $A/L\ll1$, the nonlinear term becomes negligible and therefore the linearized Navier-Stokes equation can be used. As an example, the magnitude of the thermo-mechanical fluctuations is $A\approx\sqrt{\langle x^2\rangle}\sim 0.1~\mathrm{pm}$ while the characteristic length is $L\sim1~\mathrm{\mu m}$ (see below).

The linearized Navier-Stokes equation in the frequency domain can be written as
\begin{equation}\label{eq:NavierStokes}
-\frac{i}{\delta^2}\vec{v} =-\frac{1}{\mu}\nabla P +\nabla^2\vec{v} ~.
\end{equation}
Here, $\delta=\sqrt{\mu/\rho\omega}$ is the Stokes boundary layer thickness, which is related to the frequency-based Reynold's number by $\mathrm{Re}_f=(L/\delta)^2$. Following the convention of Refs. \cite{JAP_84_64_1998,Booksection_Roukes_BioNEMS_2007}, $L=(r_{out}-r_{in})/2$ is chosen as the half-width of the micro-wheel. For the device with width of $r_{out}-r_{in}=2.5~\mathrm{\mu m}$ and frequency of 170 MHz, it gives $\delta=31$ nm and $\mathrm{Re}_f=1700$, which is very high compared with other demonstrated systems. Note that the velocity-based Reynold's number $\mathrm{Re}_u$ is still much less than one. Therefore, here we have $\mathrm{Re}_u\ll1\ll\mathrm{Re}_f$.

Consider the system shown in Fig. \ref{fig:Supp_model} (a). All mechanical motion is assumed to be rotationally symmetric, i.e., $\vec{v}(\vec{r})=v_r(r,z)\hat{r}+v_z(r,z)\hat{z}$. 
The velocity of the micro-wheel structure is given by $e^{-i\omega t} v_s\hat{r}$ while the velocity field of the surrounding fluid is given by $e^{-i\omega t} \vec{v}(r,z)$. The no-slipping condition implies that $\vec{v}(r,z)=v_s\hat{r}$ for $(r,z)\in C$ at the boundary. For systems that exhibit rotational symmetry, it is convenient to use the stream function $\psi$ to describe the fluidic motion, where $\nabla\times\vec{\psi}=\vec{v}$. Another useful quantity is the vorticity defined as $\vec{w}=\nabla\times\vec{v}$. It can be shown that both $\vec{\psi}$ and $\vec{w}$ have only azimuthal component, i.e., $\vec{\psi}=\psi(r,z)\hat{\theta}$ and $\vec{w}=w(r,z)\hat{\theta}$. By rewriting the Navier-Stokes equation in terms of $\psi$ and $w$ and applying Green's identity to convert the differential equation into integral equation, it can be shown that $\psi$ satisfies the following equation
\begin{eqnarray}\label{eq:Greens_eqn}
\frac{\psi(r^\prime,z^\prime)}{r^\prime} =\oint_C |d\vec{l}| \left[\psi\left(\partial_n G +\frac{\Phi}{r}\hat{r}\cdot\hat{n}\right) -H\partial_n\psi \right. \nonumber \\
\left.-i\delta^2\frac{P}{\mu}\partial_t\Phi +i\delta^2 w\left(\partial_n\Phi +\frac{\Phi}{r}\hat{r}\cdot\hat{n} \right) \right] ~,
\end{eqnarray}
where $\partial_n$ and $\partial_t$ represent the differentiation along the normal and transverse direction. The Green's functions $G$, $H$, $\Phi$ are given by
\begin{eqnarray}\label{eq:Greens_func}
G(r^\prime,z^\prime|r,z) &=& \frac{1}{\pi}\int_0^\infty dk \cos k(z-z^\prime) I_1(kr_<) K_1(kr_>) \nonumber \\
H(r^\prime,z^\prime|r,z) &=& \frac{1}{\pi}\int_0^\infty dk \cos k(z-z^\prime) I_1(\tilde{k}r_<) K_1(\tilde{k}r_>) \nonumber \\
\Phi(r^\prime,z^\prime|r,z) &=& G(r^\prime,z^\prime|r,z)-H(r^\prime,z^\prime|r,z) ~,
\end{eqnarray}
where $I_1(x)$ and $K_1(x)$ are the modified Bessel function of the first and second kind with index 1, $\tilde{k}=\sqrt{k^2-i/\delta^2}$, $r_<$ ($r_>$) represents $r$ or $r^\prime$ whichever is smaller (large), and $\hat{n}$ is the unit vector point outward from the surface (see Fig. \ref{fig:Supp_model} (a)). For computation of the above Green's functions with large arguments, the following asymptotic form of $I_1(x)K_1(x^\prime)$ becomes useful \cite{Book_Abramowitz_Math},
\begin{eqnarray}\label{eq:formula_IK}
I_1(x)K_1(x^\prime) \approx \frac{e^{x-x^\prime}}{2\sqrt{xx^\prime}}\left[ 1-\frac{3}{8x} +\frac{3}{8x^\prime} -\frac{9}{64xx^\prime} \right. \nonumber\\
\left. -\frac{15}{128x^2} -\frac{15}{128x^{\prime 2}} +\cdots \right] ~,
\end{eqnarray}
since it ensures that the product exponentially decays with $x$ and $x^\prime$ as long as $x^\prime>x$ and thereby avoids the overflow problem when computing $I_1$ with large arguments.

Eq. (\ref{eq:Greens_eqn}) means that the stream function $\psi$ anywhere inside the fluid is fully determined by the values of $\psi$, $\partial_n \psi$, $P$, and $w$ at the boundary $C$. Among these four functions, $\psi$ and $\partial_n \psi$ are known since they can be calculated from the velocity of the fluid at the boundary, which follows that of the mechanical resonator because of the no-slipping boundary condition. When the term on the left hand side evaluated at the boundary surface, i.e., $(r^\prime,z^\prime)\in C$, Eq. (\ref{eq:Greens_eqn}) represents a self-consistent equation with two unknown functions $P(r,z)$ and $w(r,z)$. Another independent equation can be obtained by taking gradient $\partial_{n^\prime}$ on both sides of the equation. The two integral equations with two unknown functions can be solved, for example, using the discretiziation method presented in Refs. \cite{JEngMath_3_29_1969,JAP_84_64_1998}.

After solving $P(r,z)$ and $w(r,z)$ at the boundary, the fluidic force acting on the structure surface can be calculated from the equation \cite{JEngMath_3_29_1969},
\begin{equation}\label{eq:fluid_force}
\vec{F}_{f}=\oint_C (-P\hat{n} +\mu\vec{w}\times\hat{n})dA ~.
\end{equation}
The total fluidic force is a sum of the pressure force acting perpendicular to the surface and the viscous force acting in parallel to the surface. From the fluidic force the hydrodynamic function can be obtained, i.e. $F_{f}[\omega]=m\omega^2 \Gamma[\omega]x[\omega]$. Fig. \ref{fig:Supp_model} (b) plots the calculated hydrodynamic function in the same frequency range as the measured data in Fig. \ref{fig5} (a). Water density of $\rho=1000~\mathrm{kg/m^3}$ and viscosity of $\mu=1~\mathrm{mPa\cdot s}$ are used. The device dimension are $r_{out}=10~\mathrm{\mu m}$ and $r_{in}=7.5~\mathrm{\mu m}$ in this calculation.

\end{document}